\documentclass[11pt]{article}
\input{epsf}

\usepackage{amsmath}
\usepackage{amssymb}
\usepackage{slashed}

\DeclareMathOperator{\uP}{\mathbb{P}}
\DeclareMathOperator{\bP}{\mathbb{\bar P}}
\newcommand{\NumberSet}[1]{\mathbf{#1}}

\begin{document} 
\title{Quartic amplitudes for Minkowski higher spin}       
\author{Anders K. H. Bengtsson\footnote{e-mail: anders.bengtsson@hb.se. Work supported by the Research and Education Board at the University of Bor{\aa}s.}}

\maketitle
\begin{center}School of Textiles, Engineering and Economics\break University of Bor{\aa}s\break All\' egatan 1, SE-50190 Bor\aa s, Sweden.\end{center}

\begin{abstract}
The problem of finding general quartic interaction terms between fields of higher helicities on the light-front is discussed from the point of view of calculating the corresponding amplitudes directly from the cubic vertices using BCFW recursion. Amplitude based no-go results that has appeared in the literature are reviewed and discussed and it is pointed out how they may perhaps be circumvented.
\end{abstract}

\newpage
%=============================================
\section{Introduction}\label{sec:Introduction}
%=============================================
An old and still unsolved problem in light-front higher spin gauge field theory in Minkowski space-time is the computation of general quartic interactions. The problem was addressed quite a while ago by Metsaev \cite{MetsaevQuartic1,MetsaevQuartic2} and some partial results were obtained but it seems that explicit interaction terms were not arrived at. Furthermore, the success of the Vasiliev theory of higher spin -- which is background independent but do not seem to have a consistent Minkowski limit -- has all the time cast doubts on the very existence of consistent interactions in Minkowski space-time. Adding to this list of negative indications is the computation of quartic amplitudes \cite{BenincasaCachazo2008a,BenincasaConde2011b} which yield still more restrictions on quartic interactions in Minkowski background. This paper is an attempt to review and address this problem by pointing out some still open access routes to circumventing the mounting no-go results.

Cubic self-interactions (in four spacetime dimensions) were first found in 1983 \cite{BBB1983a} and a few years later all possible cubic interactions between three fields of different spins was written down \cite{BBL1987}. These results have been generalised to higher dimensions, half-integer helicities and mixed symmetry and massive fields by Metsaev \cite{Metsaev1993a,Metsaev2005ar,Metsaev2007fb}. The spin 1 and 2 quartic interactions  are certainly known as they can be derived from the covariant theories through light-front gauge fixing. Spin 1 is rather trivial, whereas for spin 2 there is some clever algebra to perform \cite{BengtssonCederwallLindgren1983,Ananth2008}. Even the spin 3 cubic self-interaction \cite{BerendsBurgersvanDam1984} can be gauge fixed to the light-front from the BBvD covariant spin 3 interaction from 1984 \cite{IBunpubl1}. It hasn't been possible to discern any obvious pattern in the spin $1$ and $2$ light-front quartic interactions that lend itself to a generalisation to higher spin $s $ interactions. The problem of pattern recognition is also confounded by the possibility of making field redefinitions.

Anyway, we know that beyond spin 2 and cubic interactions we need a theory containing an infinite tower of higher spins in order to have any hope of finding a fully consistent interacting theory. Even at the cubic level we have interactions connecting fields of three different spin $s_1, s_2, s_3$ (with certain restrictions) \cite{BBL1987}. Therefore what is called for -- at the very least -- is a general quartic interaction term for four different spin $s_1, s_2, s_3, s_4$ (possibly with certain restrictions).

To address this problem head on using the configuration space techniques of the original 1983 paper doesn't seem to be tractable. Beyond the cubic level we  need a formalism that can maintain a spectrum of all helicities. Such a formalism becomes most natural in momentum space together with a Fock space basis in which a higher spin field $|\Phi\rangle$ can be expanded over some internal parameter, for example excitations of oscillators. Such a formalism was set up by Bengtsson et. al. \cite{BBL1987} and has been elaborated by Metsaev \cite{Metsaev1993a} and further systematised by myself \cite{AKHB2012a}. Cubic interaction data is then maintained by vertex operators that are themselves expansions in oscillators and momenta. In this way the confusing details of individual interaction terms are subsumed into the vertex operators that store the data with no particular reference to individual spin states of the interacting fields. One would hope that such vertex operators should exist also at the quartic level and above.

This said, it seems that even with the systematisation provided by a momentum space and vertex operator formalism, the quartic vertex is still elusive to calculate. This is so because there appear certain technical problems the solution of which probably requires further understanding of the cubic vertex and higher order kinematics as well as development of the formalism itself.

As already mentioned, the consistency of Minkowski spacetime higher spin theory beyond the cubics is questioned on the basis of various no-go results.  If it is the case that higher spin theory in a Minkowski background cannot be extended beyond spin 2, one would perhaps have thought that these problems should occur already at the cubic level. Instead the free theory and the cubic theory are perfectly seamless generalisations of lower spin. However, the light-front cubic interactions are only restricted by kinematics\cite{AKHB2012a} and do not really probe interactions. In terms of amplitudes this appears in the fact that the cubic amplitudes are determined by little group scaling and dimensional analysis.

A few years ago, quartic amplitudes for massless fields were investigated by Benincasa and Cachazo  \cite{BenincasaCachazo2008a} and Benincasa and Conde \cite{BenincasaConde2011a,BenincasaConde2011b}, in order to map out the consistent interactions among higher and lower spin particles. They use BCFW \cite{BCF2005,BCFW2005} recursion to build quartic tree amplitudes out of cubic amplitudes. As they show, the method -- which is a generalisation of now quite standard amplitude methods -- is very general and can be set up to test higher order consistency of gauge field theories without knowing the underlying Lagrangian. The method is based on a modern version of S-matrix theory in which five basic assumption are made: (i) analyticity, (ii) unitarity, (iii) Poincar{\'e} invariance, (iv) existence of one-particle states and (v) locality. I will try to review parts of this work below.

On the face of it, higher spin gauge interactions turn out to be very severely restricted in Minkowski spacetime. In the cited work, the focus was on single spin theories. The four-particle test (to be described in more detail below) rules out all quartic higher spin interactions except what would be generalisations of spin one $F^3$ and spin two $R^3$ interactions. 

On the other hand we know that once spin 2 is passed we need an infinite tower of higher spin excitations in order to have any hope at all at finding interactions. I will discuss how this property of higher spin field theories might escape the four-particle no-go results. There are two assumptions that can be questioned: the fifth one on locality of the S-matrix, and the fourth one on the existence of one-particle states. Of course, questioning these two basic assumption more or less amounts to questioning the S-matrix programme itself.

The first three assumptions are harder to question. Giving up Poincar{\'e} invariance is the same as bidding farewell to Minkowski higher spin which we don't want to do just as yet. 

It is customary in higher spin theory to collect an infinite spectrum (or tower) of higher spin fields into some bilocal object $\Phi(p,\xi)$ where the parameter $\xi$ works as an expansion variable over which, for instance, symmetric tensor fields can be expanded. Sometimes this is just a convenient calculational device and no particular dynamics is associated to the internal parameter. But more often $\xi$ is considered to be more of a physical variable, perhaps in connection to underlying particle mechanics constraints. In my opinion, this last viewpoint is the more physical interesting and the one more likely to lead to an eventual understanding of higher spin theory. 

Now it is -- at least heuristically -- possible to see how such a viewpoint might lead to a rejection of S-matrix assumptions (iv) and (v). A state $\Phi(p,\xi)$ depending on two variables -- a momentum $p$, and some internal coordinate $\xi$ -- is indeed the most simple non-local object one can imagine. If the expansion over $\xi$ is discrete (as in a power series) then the coefficients can be identified (at least in principle) with one-particle states. However, allowing for a more generous function space, we can (and perhaps must) drop assumption (iv) on one-particle states. Perhaps the most natural implementation of such a scenario is to connect to the continuous spin representations of the Poincar{\'e} group.

%=======================================================================
\section{The light-front and amplitudes}\label{sec:LightFrontAmplitudes}
%=======================================================================
It is interesting to discuss the light-front cubic vertices and their relation to cubic amplitudes.

%-----------------------------------------------------------------------------
\subsection{Cubic light-front vertices}\label{subsec:CubicLightFrontVertices}
Cubic interaction vertices for arbitrary helicities $\lambda_1, \lambda_2,\lambda_3$ have a remarkably simple structure when written in a light-front momentum frame \cite{BBL1987}. Taking all $\lambda$ as positive integers (denoting a negative helicity explicitly as $-\lambda$), there is essentially just one vertex for each helicity assignment. For instance a $(-\lambda_1,\lambda_2,\lambda_3)$ vertex takes the form
\begin{equation}
\frac{\gamma_1^{\lambda_1}}{\gamma_2^{\lambda_2}\gamma_3^{\lambda_3}}\bP^{\lambda_2+\lambda_3-\lambda_1}
\end{equation}
This term should be multiplied by the momentum conservation delta function and the appropriate coupling factor. In the light-front action this term is supplemented with its complex conjugate to make the action real.

In the particular case of three equal helicities $\lambda_1=\lambda_2=\lambda_3=\lambda$ we have
\begin{equation}
\left(\frac{\gamma_1}{\gamma_2\gamma_3}\bP\right)^{\lambda}
\end{equation}
This formula can be interpreted as a generalised KLT-relation \cite{KawaiLewellenTye1986} with the higher spin $\lambda$ cubic vertex being the Yang-Mills vertex raised to the power $\lambda$ as pointed out by Ananth \cite{Ananth2012un} (see also our paper \cite{BBL1987}). In these formulas, $\gamma$ is equal to the momentum component $p^+$ and $\uP$ is
\begin{equation}
\mathbb{P}=\frac{1}{3}\big((\gamma_1p_2-\gamma_2p_1)+(\gamma_2p_3-\gamma_3p_2)+(\gamma_3p_1-\gamma_1p_3)\big)\equiv\frac{1}{3}(\mathbb{P}_{12}+\mathbb{P}_{23}+\mathbb{P}_{31}) 
\end{equation} 
and correspondingly for the complex conjugate transverse momenta $\bP$. Due to momentum conservation, all three $\uP_{12}$, $\uP_{23}$ and $\uP_{31}$ are equal.

The form of the most general cubic interaction vertex on the light-front can be controlled in the following simple way, based on Poincar{\'e} analysis \cite{BBL1987, AKHB2012a}. 

Consider the cubic interaction of three fields of helicities $s_1$, $s_2$ and $s_3$ (positive or negative). As is customary we write the field as $\phi_i$ if the helicity $s_i$ is positive and $\bar{\phi}_i$ if the helicity $s_i$ is negative. The transverse momentum dependence allowed by Poincar{\'e} invariance is $\mathbb{\bar P}^n\mathbb{P}^m$ where $n$ and $m$ are restricted by $n-m=s_1+s_2+s_3$. Factors of $\mathbb{\bar P}\mathbb{P}$ correspond to field redefinitions of the free theory so we can set either $n=0$ or $m=0$. 

Treating $m=0$ first so that $n=s_1+s_2+s_3\geq0$ we get the cases
\begin{description}
  \item[(a)] All $s_i\geq0$.\; Put all $s_i=\lambda_i$. Field and momentum structure become\\

$\phi_1\phi_2\phi_3\frac{1}{\gamma_1^{\lambda_1}\gamma_2^{\lambda_2}\gamma_3^{\lambda_3}}\mathbb{\bar P}^{\lambda_1+\lambda_2+\lambda_3}$\\

  \item[(b)] Two $s_i\geq0$, one $s_i\leq0$.\; Say $s_2=\lambda_2\geq0$, $s_3=\lambda_3\geq0$ while $s_1=-\lambda_1\leq0$. Field and momentum structure become\\
 
$\bar{\phi}_1\phi_2\phi_3\frac{\gamma_1^{\lambda_1}}{\gamma_2^{\lambda_2}\gamma_3^{\lambda_3}}\mathbb{\bar P}^{\lambda_2+\lambda_3-\lambda_1}$\quad where\quad$\lambda_1\leq\lambda_2+\lambda_3$\\

\item[(c)] One $s_i\geq0$, two $s_i\leq0$.\; Say $s_3=\lambda_3\geq0$ while $s_1=-\lambda_1\leq0$, $s_2=-\lambda_2\leq0$. Field and momentum structure become\\
 
$\bar{\phi}_1\bar{\phi}_2\phi_3\frac{\gamma_1^{\lambda_1}\gamma_2^{\lambda_2}}{\gamma_3^{\lambda_3}}\mathbb{\bar P}^{\lambda_3-\lambda_1-\lambda_2}$\quad where\quad$\lambda_1+\lambda_2\leq\lambda_3$\\
\end{description}

Next treating $n=0$ so that $m=-(s_1+s_2+s_3)\geq0$ we get the cases
\begin{description}
  \item[(a')] All $s_i\leq0$.\; Put all $s_i=-\lambda_i$. Field and momentum structure become\\

$\bar{\phi}_1\bar{\phi}_2\bar{\phi}_3\frac{1}{\gamma_1^{\lambda_1}\gamma_2^{\lambda_2}\gamma_3^{\lambda_3}}\mathbb{P}^{\lambda_1+\lambda_2+\lambda_3}$\\

  \item[(b')] Two $s_i\leq0$, one $s_i\geq0$.\; Say $s_2=-\lambda_2\leq0$, $s_3=-\lambda_3\leq0$ while $s_1=\lambda_1\geq0$. Field and momentum structure become\\
 
$\phi_1\bar{\phi}_2\bar{\phi}_3\frac{\gamma_1^{\lambda_1}}{\gamma_2^{\lambda_2}\gamma_3^{\lambda_3}}\mathbb{P}^{\lambda_2+\lambda_3-\lambda_1}$\quad where\quad$\lambda_1\leq\lambda_2+\lambda_3$\\

\item[(c')] One $s_i\leq0$, two $s_i\geq0$.\; Say $s_3=-\lambda_3\leq0$ while $s_1=\lambda_1\geq0$, $s_2=\lambda_2\geq0$. Field and momentum structure become\\
 
$\phi_1\phi_2\bar{\phi}_3\frac{\gamma_1^{\lambda_1}\gamma_2^{\lambda_2}}{\gamma_3^{\lambda_3}}\mathbb{P}^{\lambda_3-\lambda_1-\lambda_2}$\quad where\quad$\lambda_1+\lambda_2\leq\lambda_3$\\
\end{description}

As expected, cases (a') - (c') are the complex conjugates of cases (a) - (c) and they can all occur in the action. Cases (a) and (a') correspond to the $F^3$ (spin $1$) and  $R^3$ (two-loop pure gravity counterterm) interactions (and their higher spin generalisations) mentioned above. 

The cases (b), (c), (b') and (c') contain, among many other types of interactions, the pure spin $s$ interactions. Just to show how it works, consider $\lambda_1=\lambda_2=\lambda_3=\lambda$. Then (b) and (b') yield the interaction term
\begin{equation}
\bar{\phi}_1\phi_2\phi_3\left(\frac{\gamma_1}{\gamma_2\gamma_3}\right)^\lambda\mathbb{\bar P}^\lambda+\phi_1\bar{\phi}_2\bar{\phi}_3\left(\frac{\gamma_1}{\gamma_2\gamma_3}\right)^\lambda\mathbb{P}^\lambda
\end{equation}
Cases (c) and (c') are not allowed for this helicity configuration.

%--------------------------------------------------------------------------
\subsection{Spinor helicity formalism}\label{subsec:SpinorHelicityFormalism}

The light-front formalism is very close to amplitude formalism as has been noted by  Ananth \cite{Ananth2012un}. Indeed, working out $\slashed{p}$ we find (with $\sigma^0$ the unit matrix)
\begin{equation}\label{eq:Pslash}
p_{a\dot a}=p_\mu\sigma^\mu_{a\dot a}=\sqrt{2}\begin{pmatrix}-p^-&\bar p\cr p &-p^+\end{pmatrix}=/\text{on shell}/=\sqrt{2}\begin{pmatrix}-\frac{p\bar p}{\gamma}&\bar p\cr p &-\gamma\end{pmatrix}
\end{equation}
Since the determinant of this two-by-two matrix is zero it can be written as a product of two-component angle and bracket spinors. Bracketing two such spinors for different momenta $p_1$ and $p_2$ we get
\begin{equation}\label{eq:AngleBracketP}
\begin{split}
\langle p_1 p_2\rangle&=\langle p_1|_{\dot a}| p_2\rangle^{\dot a}=\frac{\sqrt[4]{2}}{\sqrt{\gamma_1}}(p_1,
-\gamma_1)\frac{\sqrt[4]{2}}{\sqrt{\gamma_2}}\begin{pmatrix}-\gamma_2\cr
-p_2\end{pmatrix}=\frac{\sqrt{2}}{\sqrt{\gamma_1\gamma_2}}\mathbb{P}_{12}\\ [p_1
p_2]&=[p_1|^a|p_2]_a=-\frac{\sqrt{2}}{\sqrt{\gamma_1\gamma_2}}\mathbb{\bar{P}}_{12}
\end{split}
\end{equation}
where the translation into light-front notation is explicit. We may thus expect an almost (modulo light-front energy conservation -- see below) seamless translation between cubic amplitudes written in terms of spinor helicity variables and the light-front cubic vertices irrespective of the helicities of the external massless particles. It is of course special for the cubic interactions that the on-shell amplitude coincides up to numerical factors with the Feynman diagram vertex.

The question is now if we can exploit this dictionary between light-front variables and amplitude variables to compute quartic higher spin amplitudes directly from the cubic amplitudes. We will try to do it using BCFW recursion. But first let us spell out a little more of the translation between the formalisms. In the following the shorthand notation $|p_j\rangle=|j\rangle$ and $|p_j]=|j]$ will be used for the spinor variables. 

Going back to the formulas \eqref{eq:AngleBracketP} we should note that the left hand side inner products are $SL(2,\NumberSet{C})$ invariants, while the right hand sides are explicit light-front constructs. Let us delve into this in a little more detail. The explicit translation can be done in the following way. Start with $p_{a\dot{a}}=-\lambda_a\tilde{\lambda}_{\dot{a}}$ in {\tiny $(-+++)$} metric and introduce bra and kets according to
\begin{equation}\label{eq:spinorsbrakets}
\begin{split}
\lambda_a=|p]_a\quad&\text{with}\quad [p|^a=\epsilon^{ab}|p]_b\\
\tilde{\lambda}_{\dot a}=\langle p|_{\dot a}\quad&\text{with}\quad |p\rangle^{\dot a}=\epsilon^{\dot a\dot b}\langle p|_{\dot b}
\end{split}
\end{equation}
Then the equation $p_{a\dot{a}}=-\lambda_a\tilde{\lambda}_{\dot{a}}$ is implemented by the assignments
\begin{equation}\label{eq:LFimplementation}
\begin{split}
|p]_a=\frac{\sqrt[4]2}{\sqrt\gamma}\begin{pmatrix}\bar p\\-\gamma\end{pmatrix}\quad\text{and}\quad\langle p|_{\dot a}=\frac{\sqrt[4]2}{\sqrt\gamma}\begin{pmatrix}p\;\,&-\gamma\end{pmatrix}
\end{split}
\end{equation}
from which formulas \eqref{eq:AngleBracketP} follows. 

With the covariant angle and square bracket notation comes a set of very useful identities.
\begin{align}
\text{Anti-symmetry}:\quad& \langle p\,q\rangle=-\langle q\,p\rangle,\quad[p\,q]=-[q\,p]\label{eq:IdentitiesASym}\\
\text{Schouten identities}:\quad& \langle i\,j\rangle\langle k\,l\rangle+\langle i\,k\rangle\langle l\,j\rangle+\langle i\,l\rangle\langle j\,k\rangle=0\label{eq:IdentitiesScou1}\\
\quad&[ i\,j][ k\,l]+[ i\,k][ l\,j]+[ i\,l][ j\,k]=0\label{eq:IdentitiesScou2}\\
\text{Momentum conservation}:\quad& \sum_{i=0}^n|i\rangle[i|=0\quad\Rightarrow\quad\sum_{i=0}^n\langle p\,i\rangle[i\,q]=0\label{eq:IdentitiesMomCons}\\
\text{Squaring}:\quad& \langle p\,q\rangle[p\,q]=2p\cdot q=2(p+q)^2=-\mathbf{s}_{pq}\label{eq:IdentitiesSqua}
\end{align}
Of these equations, the anti-symmetry and Schouten identities are satisfied by the light-front form of the brackets. However, momentum conservation must be treated with some care when expressed in terms of the light-front form. Explicitly we get
\begin{equation}\label{eq:momnoncons}
\sum_{i=0}^n|i\rangle[i|=\sum_{i=0}^n\frac{\sqrt[4]2}{\sqrt\gamma_i}\begin{pmatrix}\gamma_i\\p_i\end{pmatrix}\frac{\sqrt[4]2}{\sqrt\gamma_i}\begin{pmatrix}\gamma_i\;\,&\bar p_i\end{pmatrix}=\sqrt{2}\sum_{i=0}^n\begin{pmatrix}\gamma_i&\bar p_i\\p_i&\frac{p_i\bar p_i}{\gamma_i}\end{pmatrix}
\end{equation}
The terms $\frac{p_i\bar p_i}{\gamma_i}$ do not sum to zero. Rather we have for $n=3$ 
\begin{align}\label{eq:n3noncons}
\sum_{i=0}^3\frac{p_i\bar p_i}{\gamma_i}=-\frac{\uP\bP}{\gamma_1\gamma_2\gamma_3}
\end{align}
and for $n=4$, for instance
\begin{align}\label{eq:n4noncons}
\sum_{i=0}^4\frac{p_i\bar p_i}{\gamma_i}=-\frac{\uP_{12}\bP_{23}}{\gamma_1\gamma_2\gamma_3}-\frac{\uP_{14}\bP_{43}}{\gamma_1\gamma_4\gamma_3}
\end{align}
It is a little bit interesting to see how the algebra works out. From the momentum conservation equation \eqref{eq:IdentitiesMomCons} we get, bracketing with (for instance) $\langle1|$ and $|3]$
\begin{equation*}
\begin{split}
\langle 1|\left(\,\sum_{i=0}^n|i\rangle[i|\right)|3]=\langle 12\rangle[23]+\langle 14\rangle[43]=-2\frac{\uP_{12}\bP_{23}}{\sqrt{\gamma_1\gamma_2}\sqrt{\gamma_2\gamma_3}}-2\frac{\uP_{14}\bP_{43}}{\sqrt{\gamma_1\gamma_4}\sqrt{\gamma_4\gamma_3}}
\end{split}
\end{equation*}
where we have used \eqref{eq:AngleBracketP}. On the other hand bracketing \eqref{eq:momnoncons} we get
\begin{equation*}
\begin{split}
&\frac{\sqrt[4]2}{\sqrt\gamma_1}\begin{pmatrix}p_1\;\,&-\gamma_1\end{pmatrix}
\sqrt{2}\begin{pmatrix}0&0\\0&\sum_{i=0}^4\frac{p_i\bar p_i}{\gamma_i}\end{pmatrix}
\frac{\sqrt[4]2}{\sqrt\gamma_3}\begin{pmatrix}\bar p_3\\-\gamma_3\end{pmatrix}=\\
&2\sqrt{\gamma_1\gamma_3}\sum_{i=0}^4\frac{p_i\bar p_i}{\gamma_i}=-2\frac{\uP_{12}\bP_{23}}{\sqrt{\gamma_1\gamma_2}\sqrt{\gamma_2\gamma_3}}-2\frac{\uP_{14}\bP_{43}}{\sqrt{\gamma_1\gamma_4}\sqrt{\gamma_4\gamma_3}}
\end{split}
\end{equation*}
where the equality \eqref{eq:n4noncons} is used.

The cubic formula \eqref{eq:n3noncons} is used when deriving the light-front cubic vertices. It can be expected that the corresponding quartic formula \eqref{eq:n4noncons} will play a similar role for the quartic vertex. This may be one of the technical pieces missing in earlier attempts at the full quartic vertex.

As regards the squaring equations \eqref{eq:IdentitiesSqua} we get, for instance for the Mandelstam invariant $\mathbf s$
\begin{equation}
\mathbf s=-\langle12\rangle[12]=\frac{\uP_{12}\bP_{12}}{\gamma_1\gamma_2}+\frac{\uP_{34}\bP_{34}}{\gamma_3\gamma_4}
\end{equation}
%
%----------------------------------------------------------
\subsection{Cubic amplitudes}\label{subsec:CubicAmplitudes}
We will only consider tree amplitudes and the notation $M_n[1^{s_1}2^{s_2}\ldots n^{s_n}]$ will be used for a partial (colour stripped) amplitude with $n$ outgoing on-shell massless  particles with helicities $s_1, s_2,\ldots s_n$. The $s_i$ are integers (as above, we use $\lambda_i>0$ when we want to be explicit about the sign of the helicity).

It is well-known that the cubic amplitudes are determined (up to numerical factors) by little group scaling and dimensional analysis to be either of
\begin{equation}\label{eq:CubicAmplitudes}
\begin{split}
M_3(1^{s_1} 2^{s_2}3^{s_3})&\sim[12]^{s_1+s_2-s_3}[23]^{s_2+s_3-s_1}[31]^{s_3+s_1-s_2}\quad\text{for}\;s_1+s_2+s_3>0\\
M_3(1^{s_1} 2^{s_2}3^{s_3})&\sim\langle12\rangle^{s_3-s_1-s_2}\langle23\rangle^{s_1-s_2-s_3}\langle31\rangle^{s_2-s_3-s_1}\quad\text{for}\;s_1+s_2+s_3<0
\end{split}
\end{equation} 
These amplitudes are complex conjugates of each other and for real momenta, due to momentum conservation and the on-shell conditions, they actually vanish. Indeed, $p_1+p_2+p_3=0$ and $p_i^2=0$ imply $p_i\cdot p_j=0$ for any $i$ and $j$. This gives $\langle12\rangle[12]=\langle23\rangle[23]=\langle31\rangle[31]=0$ which can only be satisfied either with all $\langle ij\rangle=0$ or all $[ij]=0$ since the spinors belong to a two-dimensional vector space.

However, for complex momenta, $\langle ij\rangle$ and $[ij]$ are independent of each other, and the cubic amplitudes in \eqref{eq:CubicAmplitudes} can be used as a basis for constructing higher order amplitudes recursively.

These amplitudes are exactly the same as the ones written above in light-front notation using the translation \eqref{eq:AngleBracketP}. This close correspondence is interesting because it could have been expected that covariantising the light-front vertices would be an awkward calculation. However, we don't know if this is special for the cubic level.

%--------------------------------------------------------------
\subsection{Coupling constants}\label{subsec:CouplingConstants}
The light-front higher spin cubic interactions discussed above should be supplied by a coupling factors $\rho(s_1,s_2,s_3)$ of mass dimension $-\vert s_1+s_2+s_3\vert$. For instance, a pure spin $s=\lambda$ interaction comes with the couplings $\rho(\lambda,\lambda,-\lambda)$ and $\rho(\lambda,-\lambda,-\lambda)$ of mass dimension $-\lambda$ as it should (to compensate the dimension of the momenta at the vertex). It is convenient to let the dimension be carried by a parameter $\rho$ of dimension $-1$ and write 
\begin{equation}\label{eq:CouplingFactor}
\rho(s_1,s_2,s_3)=\rho^{\vert s_1+s_2+s_3\vert}y_{s_1,s_2,s_3}
\end{equation}
where $y_{s_1,s_2,s_3}$ is a numerical factor. It is obvious that $y_{s_1,s_2,s_3}$ has full permutational symmetry in $s_1$, $s_2$ and $s_3$ but note that the signs of the $s$ are significant. $y_{\lambda,\lambda,\lambda}$ need not be equal to $y_{\lambda,\lambda,-\lambda}$ (for $\lambda=2$ corresponding to the cubic $R^3$ term and the spin $2$ basic cubic interaction respectively). However, we do have $y_{\lambda,\lambda,-\lambda}=y_{\lambda,-\lambda,-\lambda}$ ensuring hermiticity of the light-front action.

%======================================================================
\section{Higher spin quartic amplitudes through BCFW recursion}\label{sec:QuarticAmplitudesHigherSpin}
%======================================================================
We will attempt a computation of the general pure quartic higher spin amplitude $M_4[1^{s_1}, 2^{s_2}, 3^{s_3}, 4^{s_4}]$ using BCFW recursion closely following the procedure in Benincasa and Cachazo \cite{BenincasaCachazo2008a} specialising to the pure spin $s$ MHV amplitude with $s_1=-s_2=s_3=-s_4=s$. However, we will consider what happens when particles of all allowed helicity are summed over in the factorisation channels. It may therefore be useful to briefly discuss the BCFW method itself.

%------------------------------------------------------
\subsection{BCFW recursion}\label{subsec:BCFWRecursion}
On-shell tree level scattering amplitudes are determined by the momenta of the external particles $i$ and their types, in the case of higher spin theory, by the helicities $s_i$ encoded by polarisations $\epsilon_i$, and colour indices $\{a_ib_i\}$ for odd spin. For our purposes, it is enough to consider four-particle amplitudes with $p_1+p_2+p_3+p_4=0$ and $p_i^2=0$ for all the particles. A tree amplitude (under the assumptions of S-matrix theory) is therefore a rational function of the Lorentz invariant quantities $p_i\cdot p_j$ and $\epsilon_i\cdot p_j$ and poles can only come from internal particle propagators. For quartic amplitudes this means poles in $\mathbf{s}$, $\mathbf{t}$ and $\mathbf{u}$. There are many reviews treating amplitude methods and BCFW recursion \cite{Dixon2013Review,FengLuo2012Review,ElvangHuangBook}. Here we will just outline the procedure in the special case of four-particle amplitudes.

In practice the recursion is done by deforming two momenta, say $p_1$ and $p_2$ into $\hat{p}_1=p_1-zq$ and $\hat{p}_2=p_2+zq$ with complex $z$ and $q\cdot q=p_1\cdot q=p_2\cdot q=0$ so that momentum conservation and the on-shell conditions still hold. Then one studies the meromorphic function $M_n(z)/z$. Consider a certain internal momenta in the amplitude (that gets shifted), say $\hat{P}_k(z)=\hat{p}_1+p_3$. It goes on-shell when 
\begin{equation}\label{eq:InternalMomentumOnshell}
\hat{P}_k(z)^2=P_k^2-2zP_k\cdot q\rightarrow0\quad\Rightarrow\quad z\rightarrow z_k=\frac{P_k^2}{2P_k\cdot q}
\end{equation}
or 
\begin{equation}\label{eq:PoleZk}
z-z_k=\frac{2zP_k\cdot q-P_k^2}{2P_k\cdot q}
\end{equation}
This means that a pole in the amplitude will occur as
\begin{equation}\label{eq:PropagatorShifted}
\frac{1}{\hat{P}_k(z)^2}=\frac{1}{2P_k\cdot q}\frac{1}{z-z_k}=\frac{z_k}{P_k^2}\frac{1}{z-z_k}
\end{equation}
which -- by the way -- explains why one studies $M_n(z)/z$.

When an internal propagator goes on-shell -- corresponding to $z$ approaching the corresponding pole -- the amplitude becomes dominated by the pole and factorises into a product of two subamplitudes and the propagator. 

Consider the Riemann sphere $\mathcal{R}=\mathbb{C}\cup\{\infty\}$. If $f(z)$ is a meromorphic function on the sphere (only a finite number of singularities) then the sum of the residues at the singularities is zero.\cite{HCohnBook_note} Applying this theorem to the function $M_n^{(i,j)}(z)$ (where the superscript $(i,j)$ denotes which momenta are shifted) yields
\begin{equation}\label{eq:ResidueTheorem}
\frac{1}{2\pi i}\int_{\mathcal{R}}\frac{M_n^{(i,j)}(z)}{z}dz=M_n^{(i,j)}(0)+\sum_{k}\frac{c_k^{(i,j)}}{z_k}-\mathcal{C}_n^{(i,j)}=0
\end{equation}
%{
where $M_n^{(i,j)}(0)$ is the residue at $z=0$ to be identified by the physical amplitude. The $c_k^{(i,j)}$ are the residues at the poles $z_k$ and $\mathcal{C}_n^{(i,j)}$ is boundary term at infinity. Different deformations $(i,j)$ yield shifted amplitudes $M_n^{(i,j)}(z)$ with different locations of the poles $z_k$, all however containing the physical amplitude at $z=0$.

Consider a certain pole $z_k$ corresponding to a certain channel $z\to z_k$. As seen above, the momenta $\hat{P}_k(z)$ goes on-shell and this channel dominates. The amplitude factorises into
\begin{equation}\label{eq:Factorisation}
M_n^{(i,j)}(z)\to M_L^{(i,j)}(z_k)\frac{z_k}{P_k^2}\frac{1}{z-z_k}M_R^{(i,j)}(z_k)
\end{equation}
and the residue is given by
\begin{equation}\label{eq:ResidueFinite}
-\frac{c_k^{(i,j)}}{z_k}=M_L^{(i,j)}(z_k)\frac{1}{P_k^2}M_R^{(i,j)}(z_k)
\end{equation}
The subscripts $L$ and $R$ denote the subamplitudes, in the case of four-particle these are cubic amplitudes. The physical amplitude is given by
\begin{equation}\label{eq:PhysicalAmplitude}
M_n=M_n^{(i,j)}(0)=M_L^{(i,j)}(z_k)\frac{1}{P_k^2}M_R^{(i,j)}(z_k)+\mathcal{C}_n^{(i,j)}
\end{equation}
The boundary contribution vanishes if $\lim_{z\to\infty}M_n^{(i,j)}(z)$.

In this paper I will only consider amplitudes under the assumption that deformed amplitudes go to zero as $z\to\infty$. Benincasa and Conde\cite{BenincasaConde2011a,BenincasaConde2011b} has studied BCFW deformations with a non-zero boundary term. They show that the factorisation of the amplitudes into subamplitudes remain the same but that the propagator factor gets multiplied by a "weight" factor $f_k^{\nu,n}$ such that 
\begin{equation}\label{eq:WeightedPhysicalAmplitude}
M_n=M_L^{(i,j)}(z_k)\frac{f_k^{\nu,n}}{P_k^2}M_R^{(i,j)}(z_k)
\end{equation}
where $\nu$ denotes the power with which the boundary term diverges.

%-----------------------------------------------------------------
\subsection{Constructibility and the four-particle test}\label{subsec:ConstructibilityFourParticleTest}
Benincasa and Cachazo \cite{BenincasaCachazo2008a} introduce two concepts, \emph{constructibility} and \emph{the four-particle test}. A theory is constructible if the four-particle tree level amplitudes can completely computed from the three-particle amplitudes.

The four-particle test amounts to computing a certain amplitude using two different deformations and requiring the results to be equal. In the case of the four-particle amplitudes considered below we will deform particles $1$ and $2$ and particle $1$ and $4$ respectively using the notation $M^{(i,j)}_4(z)$ for a $i$ and $j$ deformation. Then the test amounts to requiring $M^{(1,2)}_4(0)=M^{(1,4)}_4(0)$. In practice, $M^{(1,4)}_4(0)$ is obtained from $M^{(1,2)}_4(0)$ simply by interchanging the labels $2$ and $4$ in the formula for $M^{(1,4)}_4(0)$. In terms of the kinematic invariants this means interchanging $\mathbf{s}$ and $\mathbf{t}$ (but of course polarisation factors may also change).

Benincasa and Cachazo \cite{BenincasaCachazo2008a} consider single spin theories. Here I will study what happens when one sums over an infinite spectrum of particles in the channels.

%---------------------------------------------------------
\subsection{Computing the four-particle amplitude}\label{subsec:ComputingFourParticleAmplitude}
\begin{figure}[h]\centering
\epsfbox{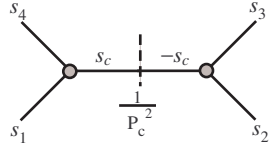}
\caption{The $\mathbf{t}$-channel contribution to the four-particle amplitude.}
\label{fig:FourAmp}
\end{figure}

In Figure \ref{fig:FourAmp} the $\mathbf{t}$-channel contribution to the amplitude we want to compute is depicted. The $\mathbf{u}$-channel should be added, but that we get by interchanging legs $3$ and $4$. The $\mathbf{s}$-channel do not contribute since we will be shifting momenta $1$ and $2$. The channel helicity is denoted by $s_c$. Eventually it will be summed over a certain range.
Writing down the parts of this amplitude (before doing the momentum shifts) using \eqref{eq:CubicAmplitudes} for the cubic amplitudes we get 
\begin{equation}\label{eq:UnshiftedQuarticExpression}
\begin{split}
\Big(&\Theta_{s_1+s_4+s_c}[14]^{s_1+s_4-s_c}[4c]^{s_4-s_1+s_c}[c1]^{s_1-s_4+s_c}+\\
&\Theta_{-(s_1+s_4+s_c)}\langle14\rangle^{s_c-s_1-s_4}\langle4c\rangle^{s_1-s_4-s_c}\langle c1\rangle^{s_4-s_1-s_c}\Big)\frac{1}{P^2_{14}}\times\\
\Big(&\Theta_{s_2+s_3-s_c}[c3]^{s_3-s_2-s_c}[32]^{s_3+s_2+s_c}[2c]^{s_2-s_3-s_c}+\\
&\Theta_{-(s_2+s_3-s_c)}\langle c3\rangle^{s_c+s_2-s_3}\langle32\rangle^{-s_c-s_3-s_2}\langle 2c\rangle^{s_3-s_2+s_c}\Big)
\end{split}
\end{equation}
where $\Theta_x$ is $1$ for $x>0$ and $0$ for $x<0$. The channel momentum $c$ is $P_{14}=p_1+p_4$. The coupling factors $\rho(s_1,s_4,s_c)$ and $\rho(s_2,s_3,s_c)$ that should multiply the $\Theta$ factors were discussed in section \ref{subsec:CouplingConstants}. They will be included when needed simply by writing for instance $\Theta_{s_1+s_4+s_c}\rightarrow\rho(s_1,s_4,s_c)\Theta_{s_1+s_4+s_c}=\rho^{\vert s_1+s_4+s_c\vert}y_{s_1,s_4,s_c}\Theta_{s_1+s_4+s_c}$.

Now we perform the following shift
\begin{equation}\label{eq:ComplexShift}
\begin{split}
|\hat1\rangle&=|1\rangle+z|2\rangle\quad\quad\quad\,|\hat1]=|1]\\
|\hat2\rangle&=|2\rangle\quad\quad\quad\quad\quad\quad|\hat2]=|2]-z|1]
\end{split}
\end{equation}
In terms of momenta this works out as
\begin{equation}\label{eq:ComplexShift}
\begin{split}
\hat{p}_1=p_1-z|1]\langle2|=p_1-z|2\rangle[1|\\
\hat{p}_2=p_2+z|1]\langle2|=p_2+z|2\rangle[1|\\
\end{split}
\end{equation}
depending on downstairs (first equality) or upstairs (second equality) dotted and un-dotted indices. Momentum conservation and the on-shell conditions are maintained since $p_1=-|1]\langle1|=-|1\rangle[1|$ and $p_2=-|2]\langle2|=-|2\rangle[2|$. It is also clear why the $\mathbf{s}$-channel momentum doesn't get shifted.

Upon performing the shift we get from \eqref{eq:UnshiftedQuarticExpression}
\begin{equation}\label{eq:ShiftedQuarticExpression}
\begin{split}
\Big(&\Theta_{s_1+s_4+s_c}[\hat{1}4]^{s_1+s_4-s_c}[4\hat{P}_{14}]^{s_4-s_1+s_c}[\hat{P}_{14}\hat{1}]^{s_1-s_4+s_c}+\\
&\Theta_{-(s_1+s_4+s_c)}\langle\hat{1}4\rangle^{s_c-s_1-s_4}\langle4\hat{P}_{14}\rangle^{s_1-s_4-s_c}\langle \hat{P}_{14}\hat{1}\rangle^{s_4-s_1-s_c}\Big)\frac{1}{\hat{P}^2_{14}}\times\\
\Big(&\Theta_{s_2+s_3-s_c}[\hat{P}_{14}3]^{s_3-s_2-s_c}[3\hat{2}]^{s_3+s_2+s_c}[\hat{2}\hat{P}_{14}]^{s_2-s_3-s_c}+\\
&\Theta_{-(s_2+s_3-s_c)}\langle \hat{P}_{14}3\rangle^{s_c+s_2-s_3}\langle3\hat{2}\rangle^{-s_c-s_3-s_2}\langle \hat{2}\hat{P}_{14}\rangle^{s_3-s_2+s_c}\Big)
\end{split}
\end{equation}
Just to be clear: this is the $\mathbf{t}$-channel part of the $M_4^{(1,2)}(z)$ amplitude. Before we are finished we shall have to add the $\mathbf{u}$-channel part (interchanging legs $3$ and $4$) and sum over channel momentum. Next it should be compared to the same amplitude computed via a $(1,4)$ shift, i.e. $M_4^{(1,4)}(z)$ (done by interchanging labels $2$ and $4$).

%§§§§§§§§§
\paragraph{Evaluation of the shifted spinors}
In order to compute the amplitude, the shifted spinors in this expression must be evaluated at the pole in $z$. From $\hat{P}_{14}^2(z)=0$ we get $\hat{P}_{14}^2=2\hat{p}_1\cdot p_4=\langle\hat{1}4\rangle[14]=0$. Therefore $z$ must be chosen such that $\langle\hat{1}4\rangle=\langle14\rangle+z\langle14\rangle=0$. Thus in the $\mathbf{t}$-channel $z_t=-\langle14\rangle/\langle24\rangle$.

For $\hat{P}_{14}(z_t)$ we get
\begin{equation}\label{eq:P14}
\hat{P}_{14}(z_t)=-|1]\langle1|+\frac{\langle14\rangle}{\langle24\rangle}|1]\langle2|-|4]\langle4|=-\frac{[14]}{[13]}|3]\langle4|
\end{equation}
using Schouten identities and momentum conservation. But we also need the spinors $|\hat{P}_{14}]$ and $\langle\hat{P}_{14}|$. From \eqref{eq:P14} we get
\begin{equation}\label{eq:P14spinors}
\begin{split}
\langle\hat{P}_{14}|&=\alpha\langle4|\\
|\hat{P}_{14}]&=\beta|3]
\end{split}
\end{equation}
with $\alpha\beta=[14]/[13]$. Furthermore, again using Schouten identities and momentum conservation to rewrite the shifted spinors of \eqref{eq:ComplexShift}, we have
\begin{equation}\label{eq:L1L2spinors}
\begin{split}
|\hat{1}\rangle&=\frac{\langle21\rangle}{\langle24\rangle}|4\rangle\quad\text{and}\quad|\hat{1}]=|1]\\
|\hat{2}]&=\frac{[12]}{[13]}|3]\quad\text{and}\quad|\hat{2}\rangle=|2\rangle
\end{split}
\end{equation}
%

%§§§§§§§§§
\paragraph{Evaluation of the subamplitudes of expression \eqref{eq:ShiftedQuarticExpression}}
\begin{description}
  \item[L1:] \begin{equation}\begin{split}&\Theta_{s_1+s_4+s_c}[\hat{1}4]^{s_1+s_4-s_c}[4\hat{P}_{14}]^{s_4-s_1+s_c}[\hat{P}_{14}\hat{1}]^{s_1-s_4+s_c}=\\&\Theta_{s_1+s_4+s_c}[14]^{s_1+s_4-s_c}[43]^{s_4-s_1+s_c}[31]^{s_1-s_4+s_c}\beta^{2s_c}\end{split}\end{equation}
  \item[L2:] \begin{equation}\begin{split}&\Theta_{-(s_1+s_4+s_c)}\langle\hat{1}4\rangle^{s_c-s_1-s_4}\langle4\hat{P}_{14}\rangle^{s_1-s_4-s_c}\langle \hat{P}_{14}\hat{1}\rangle^{s_4-s_1-s_c}=\\&\Theta_{-(s_1+s_4+s_c)}\left(\frac{\langle21\rangle}{\langle24\rangle}\right)^{-2s_1}\alpha^{-2s_c}\langle44\rangle^{-(s_1+s_4+s_c)}=0\end{split}\end{equation}
  \item[R1:] \begin{equation}\begin{split}&\Theta_{s_2+s_3-s_c}[\hat{P}_{14}3]^{s_3-s_2-s_c}[3\hat{2}]^{s_3+s_2+s_c}[\hat{2}\hat{P}_{14}]^{s_2-s_3-s_c}=\\&\Theta_{s_2+s_3-s_c}\left(\frac{[12]}{[13]}\right)^{2s_2}\beta^{-2s_2}[33]^{s_2+s_3-s_c}=0\end{split}\end{equation}
  \item[R2:] \begin{equation}\begin{split}&\Theta_{-(s_2+s_3-s_c)}\langle \hat{P}_{14}3\rangle^{s_c+s_2-s_3}\langle3\hat{2}\rangle^{-s_c-s_3-s_2}\langle \hat{2}\hat{P}_{14}\rangle^{s_3-s_2+s_c}=\\&\Theta_{-(s_2+s_3-s_c)}\langle43\rangle^{s_c+s_2-s_3}\langle32\rangle^{-s_c-s_3-s_2}\langle24\rangle^{s_3-s_2+s_c}\alpha^{2s_c}\end{split}\end{equation}
\end{description} 

We can now put together the $\mathbf{t}$-channel contribution to $M_4^{(1,2)}(0)$. After some algebra we get\cite{BenincasaCachazo2008a}
\begin{equation}\label{eq:TchannelM4}
\begin{split}
&M_{4,\mathbf{t}}^{(1,2)}(0)=\Theta_{s_1+s_4+s_c}\Theta_{-(s_2+s_3-s_c)}\frac{(-P_{3,4}^2)^{s_c}}{P_{1,4}^2}\times\\&\left(\frac{[14][31]}{[43]}\right)^{s_1}\left(\frac{\langle43\rangle}{\langle32\rangle\langle24\rangle}\right)^{s_2}\left(\frac{\langle24\rangle}{\langle43\rangle\langle32\rangle}\right)^{s_3}\left(\frac{[14][43]}{[31]}\right)^{s_4}
\end{split}
\end{equation}
These terms contribute to the channel sum over $s_c$ for $s_c>\max(-(s_1+s_4),(s_2+s_3))$. The $\mathbf{u}$-channel contribution to $M_4^{(1,2)}(0)$ is obtained by interchanging labels $3$ and $4$.
\begin{equation}\label{eq:UchannelM4}
\begin{split}
&M_{4,\mathbf{u}}^{(1,2)}(0)=\Theta_{s_1+s_3+s_c}\Theta_{-(s_2+s_4-s_c)}\frac{(-P_{4,3}^2)^{s_c}}{P_{1,3}^2}\times\\&\left(\frac{[13][41]}{[34]}\right)^{s_1}\left(\frac{\langle34\rangle}{\langle42\rangle\langle23\rangle}\right)^{s_2}\left(\frac{\langle23\rangle}{\langle34\rangle\langle42\rangle}\right)^{s_4}\left(\frac{[13][34]}{[41]}\right)^{s_3}
\end{split}
\end{equation}
These terms contribute to the channel sum over $s_c$ for $s_c>\max(-(s_1+s_3),(s_2+s_4))$.

%---------------------------------------------------------
\subsection{Pure spin $s$ four-particle MHV amplitude}\label{subsec:PureSpinSFourParticleMHVAmplitude}
We now have enough data to compute a pure spin $s$ MHV four-particle amplitude with $s_1=-s_2=s_3=-s_4=s$. Then the channel sum runs $s_c>0$ in the $\mathbf{t}$-channel contribution and $s_c>-2s$ $\mathbf{u}$-channel contribution. Also introduce the coupling factors according to \eqref{eq:CouplingFactor}
\begin{equation}
\begin{split}
\rho(s_1,s_4,s_c)\rho(s_2,s_3,-s_c)&=\rho(s,-s,s_c)\rho(-s,s,-s_c)\\
&=\rho^{2|s_c|}y_{s,-s,s_c}y_{-s,s,-s_c}=\rho^{2|s_c|}y_{s,-s,s_c}^2
\end{split}
\end{equation}
for the $\mathbf{t}$-channel, and
\begin{equation}
\begin{split}
\rho(s_1,s_3,s_c)\rho(s_2,s_4,-s_c)&=\rho(s,s,s_c)\rho(-s,-s,-s_c)\\
&=\rho^{2|s_c+2s|}y_{s,s,s_c}y_{-s,-s,-s_c}=\rho^{2|s_c+2s|}y_{s,s,s_c}^2
\end{split}
\end{equation}

Specialising formulas \eqref{eq:TchannelM4} and \eqref{eq:UchannelM4} and submitting them to some further algebra we get for the $\mathbf{t}$-channel
\begin{equation}\label{eq:TchannelSum}
-\sum_{s_c>0}\rho^{2|s_c|}y_{s,-s,s_c}^2\frac{\mathbf{s}^{s_c-2s}}{\mathbf{t}}\left([31]\langle24\rangle\right)^{2s}
\end{equation}
and for the $\mathbf{u}$-channel
\begin{equation}\label{eq:UchannelSum}
-\sum_{s_c>-2s}\rho^{2|s_c+2s|}y_{s,s,s_c}^2\frac{\mathbf{s}^{s_c}}{\mathbf{u}}\left([31]\langle24\rangle\right)^{2s}
\end{equation}
The sum of these two sums is a candidate for a pure spin $s$ four-particle MHV amplitude computed by a $(1,2)$ BCFW shift. For a further simplification, let us only consider even spin (so that we need not take care of colour factors), for concreteness scattering of spin $s=2$ and $s=4$ respectively.

%§§§§§§§§§
\paragraph{Pure spin 2 scattering}
\begin{equation}\label{eq:PureSpin2}
\begin{split}
-\left([31]\langle24\rangle\right)^{4}\Big(&\rho^4y_{2,-2,2}^2\frac{\mathbf{s}^{-2}}{\mathbf{t}}+\rho^8y_{2,-2,4}^2\frac{\mathbf{s}^{0}}{\mathbf{t}}+\rho^{12}y_{2,-2,6}^2\frac{\mathbf{s}^{2}}{\mathbf{t}}+\ldots\\
&\rho^4y_{2,2,-2}^2\frac{\mathbf{s}^{-2}}{\mathbf{u}}+\rho^8y_{2,2,0}^2\frac{\mathbf{s}^{0}}{\mathbf{u}}+\rho^{12}y_{2,2,2}^2\frac{\mathbf{s}^{2}}{\mathbf{u}}+\ldots\Big)
\end{split}
\end{equation}
A theory with only spin $2$ in the channel would have only $y_{2,-2,2}=y_{2,2,-2}$ non-zero with an amplitude depending on the kinematic invariants as
\begin{equation}\label{eq:PureSpin2Channel2}
\sim\mathbf{s}^{-2}\left(\frac{1}{\mathbf{t}}+\frac{1}{\mathbf{u}}\right)=\frac{1}{\mathbf{s}\mathbf{t}\mathbf{u}}
\end{equation}
The amplitude computed using a $(1,4)$ shift can be obtained by interchanging labels $2$ and $4$ (as discussed above). The prefactor $[31]\langle24\rangle$ and $\mathbf{u}$ stays unchanged while $\mathbf{s}\leftrightarrow\mathbf{t}$. The form of the prefactor depends on the helicity configuration $s_1=s_3=-s_2=-s_4=s$. Therefore the amplitudes that results from a $(1,2)$ shift and a $(1,4)$ shift are the same. But note that one is then actually discarding a pure spin $2$ term in the $\mathbf{u}$ channel, namely 
\begin{equation*}
\rho^{12}y_{2,2,2}^2\frac{\mathbf{s}^{2}}{\mathbf{u}}
\end{equation*}
corresponding to the (cubic level) $R^3$ two-loop counterterm.

%§§§§§§§§§
\paragraph{Pure spin 4 scattering}
\begin{equation}\label{eq:PureSpin4}
\begin{split}
-\left([31]\langle24\rangle\right)^{8}\Big(&\rho^4y_{4,-4,2}^2\frac{\mathbf{s}^{-6}}{\mathbf{t}}+\rho^8y_{4,-4,4}^2\frac{\mathbf{s}^{-4}}{\mathbf{t}}+\rho^{12}y_{4,-4,6}^2\frac{\mathbf{s}^{-2}}{\mathbf{t}}+\ldots\\
&\rho^4y_{4,4,-6}^2\frac{\mathbf{s}^{-6}}{\mathbf{u}}+\rho^8y_{4,4,-4}^2\frac{\mathbf{s}^{-4}}{\mathbf{u}}+\rho^{12}y_{4,4,-2}^2\frac{\mathbf{s}^{-2}}{\mathbf{u}}+\ldots\Big)
\end{split}
\end{equation}
A theory with only spin $4$ in the channel and just one coupling constant $\rho^4y_{4,4,-4}$ (and thus discardaring a pure spin $4$ term further up in the $\mathbf{u}$-channel) would have a dependence on the kinematic invariants
\begin{equation}\label{eq:PureSpin4Channel4_12}
\sim\mathbf{s}^{-4}\left(\frac{1}{\mathbf{t}}+\frac{1}{\mathbf{u}}\right)=\frac{\mathbf{s}^{-2}}{\mathbf{s}\mathbf{t}\mathbf{u}}
\end{equation}
Again, the amplitude computed using a $(1,4)$ shift can be obtained by interchanging labels $2$ and $4$, with the result for the dependence on kinematic invariants
\begin{equation}\label{eq:PureSpin4Channel4_14}
\sim\mathbf{t}^{-4}\left(\frac{1}{\mathbf{s}}+\frac{1}{\mathbf{u}}\right)=\frac{\mathbf{t}^{-2}}{\mathbf{s}\mathbf{t}\mathbf{u}}
\end{equation}
So, according to the four-particle test, pure spin $4$ scattering is ruled out since $M_4^{(1,2)}\neq M_4^{(1,4)}$. Let us look at the general pure spin $s$ scattering.

%§§§§§§§§§
\paragraph{Pure even spin $s$ scattering}
\begin{equation}\label{eq:PureSpin4}
\begin{split}
-\left([31]\langle24\rangle\right)^{2s}\Big(&\rho^4y_{s,-s,2}^2\frac{\mathbf{s}^{2-2s}}{\mathbf{t}}+\rho^8y_{s,-s,4}^2\frac{\mathbf{s}^{4-2s}}{\mathbf{t}}+\rho^{12}y_{s,-s,6}^2\frac{\mathbf{s}^{6-2s}}{\mathbf{t}}+\ldots\\
&\rho^4y_{s,s,2-2s}^2\frac{\mathbf{s}^{2-2s}}{\mathbf{u}}+\rho^8y_{s,s,4-2s}^2\frac{\mathbf{s}^{4-2s}}{\mathbf{u}}+\rho^{12}y_{s,s,6-2s}^2\frac{\mathbf{s}^{6-2s}}{\mathbf{u}}+\ldots\Big)
\end{split}
\end{equation}
Picking out the terms corresponding to spin $s$ in the channel and uniform dimension of the coupling constant, we get for the $(1,2)$ shift
\begin{equation}
M_4^{(1,2)}\sim\left([31]\langle24\rangle\right)^{2s}\rho^{2s}y_{s,s,-s}^2\mathbf{s}^{-s}\left(\frac{1}{\mathbf{t}}+\frac{1}{\mathbf{u}}\right)\sim\frac{\mathbf{s}^{2-s}}{\mathbf{s}\mathbf{t}\mathbf{u}}
\end{equation}
and for the $(1,4)$ shift
\begin{equation}
M_4^{(1,4)}\sim\left([31]\langle24\rangle\right)^{2s}\rho^{2s}y_{s,s,-s}^2\mathbf{t}^{-s}\left(\frac{1}{\mathbf{s}}+\frac{1}{\mathbf{u}}\right)\sim\frac{\mathbf{t}^{2-s}}{\mathbf{s}\mathbf{t}\mathbf{u}}
\end{equation}
This essentially the four-particle test as derived by Benincasa and Cachazo\cite{BenincasaCachazo2008a}. Demanding the two expressions for the scattering amplitude to be the same forces $s=2$, ruling out higher spin scattering.

%--------------------------------------------------
\subsection{End of story?}\label{subsec:EndStory}
Embarrasing as this result is for Minkowski higher spin, it is not yet the end of story. First of all, we really have to study the full series of terms i \eqref{eq:TchannelSum} and  \eqref{eq:UchannelSum}. A somewhat similar analysis has been reported by Dempster and Tsulaia \cite{DempsterTsulaia2012}. 

For concreteness, study the case of pure spin $2$ scattering. Let us also assume that the numerical $y$ coefficients are equal at each power $\mathbf{s}$: $y_{2,-2,n}=y_{2,-2,n-4}\equiv y_n$. This amounts to, for instance, that the spin $(2,2,2)$ coupling comes with the same numerical strength as the spin $(2,-2,6)$ coupling involving a spin $6$ field (which of course need not be the case). We get

\begin{equation}\label{eq:PureSpin2Summed_12}
\begin{split}
-\left([31]\langle24\rangle\right)^{4}\Big(&\rho^4y_{2,-2,2}^2\frac{\mathbf{s}^{-2}}{\mathbf{t}}+\rho^8y_{2,-2,4}^2\frac{\mathbf{s}^{0}}{\mathbf{t}}+\rho^{12}y_{2,-2,6}^2\frac{\mathbf{s}^{2}}{\mathbf{t}}+\ldots\\
&\rho^4y_{2,2,-2}^2\frac{\mathbf{s}^{-2}}{\mathbf{u}}+\rho^8y_{2,2,0}^2\frac{\mathbf{s}^{0}}{\mathbf{u}}+\rho^{12}y_{2,2,2}^2\frac{\mathbf{s}^{2}}{\mathbf{u}}+\ldots\Big)=\\
&-\left([31]\langle24\rangle\right)^{4}\rho^4\frac{1}{\mathbf{s}\mathbf{t}\mathbf{u}}\big(y_2^2+y_4^2\rho^4\mathbf{s}^2+y_6^2\rho^8\mathbf{s}^4+\ldots\big)
\end{split}
\end{equation}
The corresponding expression coming from the $(1,4)$ shift becomes
\begin{equation}\label{eq:PureSpin2Summed_14}
-\left([31]\langle24\rangle\right)^{4}\rho^4\frac{1}{\mathbf{s}\mathbf{t}\mathbf{u}}\big(y_2^2+y_4^2\rho^4\mathbf{t}^2+y_6^2\rho^8\mathbf{t}^4+\ldots\big)
\end{equation}
The expressions still disagree. Considered as power series they will have a certain radius of convergence within which they certainly differ and outside of which they both diverge. However, both results make some sense. We know that higher spin theory entails ever increasing powers of momenta in the interactions, cubic and higher order. In terms of amplitudes this most likely would manifest itself through increasing powers of kinematic invariants.

We see that computing the amplitude through a $(1,2)$ shift, we pick up poles in $\mathbf{t}$ and $\mathbf{u}$ while we get power series in the $\mathbf{s}$ channel. Likewise, computing the amplitude through a $(1,4)$ shift, we pick up poles in $\mathbf{s}$ and $\mathbf{u}$ while we get power series in the $\mathbf{t}$ channel. In both cases the missing pole also appears when adding the two series.

Had we computed the amplitude through a $(1,3)$ shift, we would get poles in $\mathbf{s}$, $\mathbf{t}$ and $\mathbf{u}$ multiplying a power series in the $\mathbf{u}$ channel. Indeed, we would get by interchanging labels $2$ and $3$
\begin{equation}\label{eq:PureSpin2Summed_13}
-\left([21]\langle34\rangle\right)^{4}\rho^4\frac{1}{\mathbf{s}\mathbf{t}\mathbf{u}}\big(y_2^2+y_4^2\rho^4\mathbf{u}^2+y_6^2\rho^8\mathbf{u}^4+\ldots\big)
\end{equation}
where the polarisation factor changes since the helicity configuration changes.

These circumstances throw some doubt on the combined power och constructibility and the four-particle test. Even though quartic amplitudes for lower spin theories can be fully reconstructed by BCFW recursion using just one type of shift, it is not clear to me that it should be sufficient for higher spin. Perhaps a certain shift only probes part of the full amplitude and that rather than requiring equality of the results of different shifts, perhaps we should combine them. The full amplitude then shows poles in all channels as well as power series in all invariants. However, this line of thought must be supplemented by a careful analysis of the $z\to\infty$ limit of the terms in the shifted amplitudes and the behaviour of the boundary term. This analysis remains to be done.

Furthermore, it may be that higher spin scattering amplitudes are not constructible, meaning that there could be an irreducible part to the four-particle amplitude that has to be added to the terms coming from cubic amplitudes. Which seems to return the question to computing the full quartic (Feynman) vertex.

%======================================================================
\section{Can Minkowski higher spin be rescued?}\label{sec:ProspectsProblems}
%======================================================================
The answer to this question --  if there is a positive one -- must be sought in physical understanding as much as in technical calculations. There is one basic physical puzzle in all of higher spin theory. If there are higher spin excitations in nature: then where are they, what do they do and what role do they play? 

That we are thinking about extreme high energy phenomena is obvious. Perhaps one could speak of ''beyond quantum gravity'' which is probably what people have in mind when they write about the subject in terms of ''higher spin gravity''. With this in mind, what sense does it make to think of scattering spin $s$ particles on spin $s^\prime$ particles? Not very much it seems.

Exciting one higher spin particle must clearly excite a whole bunch of them -- indeed all of them -- if we take seriously what we already know. So, in my opinion, considering particular higher spin fields by themselves doesn't make much sense. What we should consider is the full spectrum of fields as one object. One should work with non-local objects such as $\Phi(x,\xi)$. I have no idea of how an S-matrix theory of such an object would look like. But it should be possible to -- formally at least -- join all cubic subamplitudes into one cubic amplitude maintaining them all. One way of doing that in a principled way could be to consider the component higher spin fields as components of a continuous spin field. The continuous spin representations naturally organize all helicities as well as a naturally involve a dimensionfull parameter. On a computational level that would mean not only summing over channel spin but also over external spin $s_1,s_2,s_3,s_4$ corresponding to scattering continuous spin fields on contiuous spin fields. This is clearly a little bit speculative, but not more so than that a calculation along these lines could be attempted. 

%======================================================================
\section*{Acknowledgement}\label{sec:Acknowledgement}
%======================================================================
This paper is an extended and modified version of a presentation I gave at the international workshop on Higher Spin Gauge Theories at the Institute of Advanced Studies at Nanyang Technological University in Singapore, November 4 - 6, 2015. I'd like to thank the organizers for the invititation and the opportunity it gave to look into this fascinating aspect of higher spin theory (of which I wasn't up to date when giving my talk). I also thank Slava Didenko for pointing out the work by Metsaev on quartic interactions that had slipped my memory.

%\begin{thebibliography}{9} % for non BIBTeX users
%\bibitem{ams04} \AmS, \emph{\AmS-\LaTeX{} ...
%\end{thebibliography}
\bibliographystyle{ws-rv-van}

%\bibliography{C:/Users/ABE/Documents/TexBibFiles/HepBibDB}
%\printindex[aindx] % to print author index
%\printindex

\end{document}